\documentclass[fleqn,twoside]{article}
\usepackage{espcrc2}
\usepackage{graphicx}

\newcommand{\AmS}{{\protect\the\textfont2
  A\kern-.1667em\lower.5ex\hbox{M}\kern-.125emS}}
   \newcommand{\beq}{\begin{equation}} 
   \newcommand{\eeq}{\end{equation}}
\newcommand{\beqa}{\begin{eqnarray}} 
   \newcommand{\eeqa}{\end{eqnarray}}   
   \def\esim{\mathrel{\rlap{\raise2pt\hbox{$\sim$}}
    \lower1pt\hbox{$-$}}}         

\def\gsim{\mathrel{\rlap{\lower4pt\hbox{\hskip1pt$\sim$}}
    \raise1pt\hbox{$>$}}}         

\title{Dark Matter Constituents}

\author{L. Bergstr\"om\address[MCSD]{Physics Department, Stockholm University, \\
AlbaNova University Centre, S-106 91  Stockholm, Sweden}
\thanks{Supported by the Swedish Research Council (VR)}}
       
\begin{document}

\begin{abstract}
As cosmology has entered a phase of precision experiments, the content of
the universe has been established to contain interesting and not yet fully 
understood components,
namely dark energy and dark matter. While the cause and exact nature of the 
dark energy remains mysterious, there is greater hope to connect the dark 
matter to current models of particle physics.
Supersymmetric models  
provide several excellent candidates for dark matter, with the lightest
neutralino the prime example. This and other dark matter candidates 
are discussed, and prospects for their detection
summarized. Some methods of detection are explained, and 
indications of signals in present data are critically examined. 
\end{abstract}


\maketitle

\section{INTRODUCTION}

The question of the nature of the average composition of the Universe is 
getting more interesting than ever. As new observations have arrived, 
the picture emerging has become quite 
puzzling, with the favoured model now containing both dark energy 
(vacuum energy, ``cosmological constant'' or perhaps a slowly evolving
scalar field, ``quintessence'') and dark, non-baryonic matter.

The recent supernova discovery of acceleration of the cosmic expansion \cite{perlmutter}
taken in combination with the spectacular new WMAP microwave background measurements \cite{wmap}
has led to a  ``concordande model''
($\Lambda$CDM): 
\beq
\left\{
\begin{array}{ccc}
        \Omega_M & = & 0.3  \\
        \Omega_\Lambda & = & 0.7\label{eq:eq1}
\end{array}\right.
\eeq
This is in nice agreement with the inflationary prediction of the matter
and cosmological constant contributions adding up to unity,
$\Omega_{tot}=1$, but produces the mystery of having a non-zero
but exceedingly small value of the cosmological constant $\Lambda$ (compared to the natural
scale of gravity, which is the Planck scale). However, the supernova
data (of two independent groups) and CMBR measurement in 
combination with the measured range of the Hubble constant ($H_0\simeq 70$
km/s\,Mpc$^{-1}$, $h\equiv H_0/100 =0.7\pm 0.1$)
has more or less forced this model upon us.

The WMAP data and Big Bang nucleosynthesis (BBN) deuterium data \cite{tytler}
 agree on
 the  baryonic contribution  $\Omega_{b}$ 
to $\Omega_M$:
\beq
\Omega_{b}h^2\esim 0.023\pm 0.002.\label{eq:bbn}
\eeq
This means that non-baryonic dark matter is required beyond any doubt, 
since $h=0.7\pm 0.1$ gives the baryonic 
contribution $\Omega_b$ less than around $0.08$, so much less than the
measured $\Omega_M$.

Weakly interacting massive particle candidates (WIMPs) for dark matter are generically found to decouple at a 
temperature 
 that is roughly $m_{\rm WIMP}/20$, which means that they
 are non-relativistic and behave as CDM 
when structure forms. 

WIMPs have the advantage that their relic density is determined by
the calculable departure from thermal equilibrium in the early universe,
and give without difficulties a contribution to $\Omega$ which may be
close to the observed value. 

\section{\uppercase{ Dark Matter Candidates}}
\subsection{Neutrinos}
With the masses of neutrinos in view of observed oscillations 
being definitely non-zero,
 we may for the first 
time say with certainty that non-baryonic dark matter exists. However,
the neutrino contribution to $\Omega_M$ is most likely very small.
For small mass neutrinos (hot dark matter, HDM), the relic abundance is given 
by
\beq
\Omega_\nu h^2={\sum_i m_{\nu_i}\over 94\ {\rm eV}}.\label{eq:relicneutrino}
\eeq

For the kind of mass values indicated by the Super-Kamiokande measurements,
0.01 -- 0.1 eV (unless there are unnatural neutrino mass degeneracies), we see that although
neutrinos indeed contribute to non-baryonic dark matter, they do so at
the level or $10^{-2}$ or less. This may seem tiny from the cosmological point
of view, but is still about as much as all visible material like stars, gas,
dust, etc contribute to the energy density of the Universe. 

The WMAP results, especially when augmented by large galaxy survey 
data such as SDSS \cite{sdss} or 2DF \cite{2df}, also give very little
room for the standard neutrino contribution to dark matter.
Since the
distribution of matter is quite different in CDM and HDM 
models (with the latter lacking small-scale power due to free-streaming), 
one may use the observed distribution to put a direct upper bound to
the sum of all neutrino masses which turns out to be of the order of
an eV or so (the WMAP limit is 0.23 eV for all three neutrinos degenerate,
but this includes less understood smaller scale data from the 
Lyman alpha forest; for a discussion see also \cite{steen}).  The advantage
of the cosmological limit is that it is sensitive to the absolute mass scale 
of neutrinos whereas neutrino oscillations only probe mass
differences.

In principle, one could imagine having a sterile neutrino as 
dark matter, if it is non-thermally produced, e.g., 
generated through mixing with the active
neutrinos. Generally, this candidate will have to have a  mass in the keV to
MeV range and would act as something inbetween cold and hot dark
matter (sometimes named ``warm dark matter'', WDM). An unpleasant feature 
of these models is a necessary, delicate finetuning of the mixing angle
versus mass to get the right abundance, but models of this kind 
have been constructed which
may perhaps evade experimental constraints \cite{dolgovhansen,abazajian}.

 \subsection{Other Solutions to the Dark Matter Problem}
 
A class of dark matter candidates, which should be kept in mind,
are
super-heavy relics which were produced by some non-thermal mechanism
in the early Universe. Examples are given by 
 ``wimpzillas'' \cite{wimpzillas}. 

At the other end of the mass spectrum are axions, light pseudoscalar 
mesons that  appear in all string theories and which are needed
to explain the weakness of CP violation in strong interactions. There
are experiments which search for these particles in mass ranges interesting 
for cosmology \cite{axion}. 

There has recently been interest in dark matter in the form of topological,
extended objects, ``Q-balls'' which may or may not be supersymmetric and 
may or may not be stable on cosmological time scales (see \cite{kari} for 
a recent review). In certain scenarios,
supersymmetric Q-balls may decay into the lightest neutralino, giving
a similar scenario as in the MSSM, but with different phenomenology. For 
instance, the usual bounds on the relic density for thermally produced
neutralinos may be evaded. 

\section{\uppercase{ Supersymmetric Dark Matter}}
One of the prime 
candidates for the non-baryonic component is provided by the lightest 
supersymmetric particle, plausibly the lightest neutralino $\chi$.

The lightest
neutralino $\chi$ is a mixture 
of the supersymmetric partners of the photon, the $Z$ and the neutral part of 
the two 
Higgs doublets present in the minimal extension of the 
supersymmetric standard model (for reviews see, e.g., \cite{jkg,lbreview}). 
The 
attractiveness
of this dark matter candidate stems from the fact that its generic couplings
 and mass range
naturally gives a relic density in the required range to explain halo 
dark matter. Besides, its 
motivation from particle physics, which was originally based on solving the
so-called hierarchy problem (the puzzling discrepancy between the
mass scales of electroweak interactions and gravity), has become stronger due to 
the apparent need for 100 GeV - 10 TeV scale supersymmetry to achieve
unification of the gauge couplings in view of LEP results \cite{amaldi}, and the prediction that the lightest Higgs boson
should be below 135 GeV, as may be favoured by LEP data \cite{higgslimit}.

Recently, there has been  discussion 
about the constraints on the
MSSM which follow from the measurements of $(g-2)_\mu$, the anomalous 
magnetic moment of the muon \cite{gminus2data}.
The new set 
of data  recently released shows a mild discrepancy with the Standard Model (but only
at
the 2 -- 3 $\sigma$ level), which could in principle be due to supersymmetry \cite{tedpaolo}.

The relic density calculation in the MSSM for a given set of 
parameters is nowadays accurate a few \% or so, thus matching the
precision given by cosmological measurements. A recent 
important improvement is the inclusion of 
coannihilations, which can change the relic abundance by a large 
factor in some instances \cite{coann}. Much of the effort that has gone
into this field has resulted in publicly available computer program 
packages, for instance DarkSUSY \cite{darksusy}.

\subsection{Detection methods}
\subsubsection{Direct detection}
If neutralinos are indeed the CDM needed on galaxy scales and larger, 
there should be a substantial flux of these particles in the Milky 
Way halo. Since the interaction strength  is 
essentially given by the same weak couplings as, e.g., for neutrinos 
there is a non-negligible chance of detecting them in low-background 
counting experiments. Due to the large parameter space of MSSM, even 
with the simplifying assumptions above, there is a rather wide span of 
predictions for the event rate in detectors of various types. 

The experimental situation is becoming interesting as several
direct detection experiments after many years of continuing
sophistication are starting to probe interesting parts of
the parameter space of the MSSM, given reasonable, central values
of the astrophysical and nuclear physics parameters. 

The most enigmatic finding is that of DAMA \cite{dama}, 
which after seven years 
of running in the low-background environment in Gran Sasso, now reports
a visible seasonal variation \cite{freese} 
of the scattering rate which is formally
at the $6\sigma$ level.

This seems to be difficult to reconcile with results from the CDMS \cite{cdms}
and EDELWEISS experiments \cite{edelweiss}, 
which have recently improved their limits by
an order of magnitude. The data released essentially
exclude all the parameter range for supersymmetric dark matter
indicated by the DAMA experiment \cite{dama}. 
However, it is difficult to compare the experiments directly, since they 
use different materials and very different techniques. For instance, it
is not excluded that a large spin-dependent interaction may explain
the DAMA results \cite{marc2}.

Although  neutralino dark matter may, with careful choice of physical
and astrophysical parameters \cite{sandro}, explain the DAMA results,
the rates in minimal supersymmetry are generally lower. Even the CDMS and 
EDELWEISS experiments are so far only scratching the surface of the
parameter space that is provided by the MSSM. However, a new generation
of experiments is being prepared with an order of magnitude better
sensitivity \cite{morales}. The urgent challenge to the experimental
dark matter community is now to build an independent, and preferably more
sensitive detector which is similar enough to DAMA to enable a direct
comparison \cite{bahcall}.
\subsection{Indirect detection}

There is also the possibility of indirect detection through 
 neutralino annihilation 
in the galactic halo or in astrophysical objects like the Sun 
and the Earth. This is becoming a promising method thanks
 to very powerful new detectors for 
cosmic gamma rays  and neutrinos planned and under construction.
 
There has recently been  balloon-borne  detection experiments \cite{HEAT}, 
where an excess of positrons over that expected from ordinary sources 
has been found. However, since there are many other possibilities to 
create positrons by astrophysical sources, e.g. near the centre of 
the Milky Way, the interpretation is not yet conclusive. In particular,
it has been difficult to reproduce the apparent ``bump'' seen at 7 -- 8
GeV in any model based on WIMP annihilation (for recent attempts, see \cite{positrons}).

 Antiprotons, $\bar p$,
from neutralino annihilations were long hoped to give a useful
signal \cite{antiprotons}, and there have been several balloon-borne
experiments \cite{caprice,bess} performed
and a very ambitious space experiment, AMS,
to search for antimatter is under way \cite{ting}.
For kinematical reasons, antiprotons
created by pair-production in cosmic ray collisions with interstellar
gas and dust are born with relatively high energy, whereas
antiprotons from neutralino annihilation populate also the sub-100
MeV energy band.

However, it was found \cite{pbar,tomg} that the cosmic-ray
induced antiprotons may populate also the low-energy region
to a greater extent than previously thought, making the extraction
of an eventual supersymmetric signal much more difficult. 
Recently, a thorough reanalysis \cite{donato03} has arrived at
the same conclusion.

Another
 problem that plagues estimates of the signal strength of both positrons and
antiprotons is  the uncertainty of the galactic propagation model
and solar wind
 modulation.

 Even allowing for large such systematic effects, the
 measured antiproton flux gives, however, rather stringent limits on
MSSM models with the highest annihilation rates. One can also
use the experimental upper limits to bound from below the
 lifetime of hypothetical $R$-parity violating decaying neutralinos \cite{baltz}. 
There may in some scenarios with a clumpy halo \cite{clumpy}
(which enhances the annihilation rate) be a possibility to detect
heavy neutralinos through spectral features above several GeV \cite{piero},
something which will be probed by the upcoming PAMELA satellite project \cite{pamela}.

A very rare process in proton-proton collisions, antideuteron production,
may be less rare in neutralino annihilation \cite{fiorenza}. However,
the fluxes are so small that the possibility of detection seems marginal
even in the AMS experiment.


With these problems of positrons and antiprotons, one would expect that 
problems
 of gamma rays and neutrinos are similar, if they arise from 
secondary decays in the 
annihilation process in the halo. For instance, the gamma ray spectrum arising from
the fragmentation of fermion and gauge boson final states is quite 
featureless and gives the bulk of the gammas at low energy where the
cosmic gamma ray background is severe. Also, the density of 
neutralinos in the halo is not large enough to give a measurable flux 
of secondary neutrinos, unless the dark matter halo is very clumpy \cite{clumpy}.
 However, neutrinos can escape from the centre  
of the Sun or Earth, where 
neutralinos may have been gravitationally trapped and therefore their density 
enhanced. 

Also, gamma-rays may be emitted from places where the dark matter density
is substantially higher than average, e.g. in dark matter clumps \cite{clumpy},
or near the galactic center \cite{bertone}. For a recent attempt to 
interpret the various anomalies in the measured fluxes of gamma rays, 
positrons and antiprotons in terms of neutralino dark matter, 
see \cite{deboer}. As a general comment, one should observe that the 
systematic uncertainties that come from unknown astrophysical parameters and
propagation functions are much larger than the statistical errors used
in \cite{deboer}.

\subsection{Gamma ray lines}

Gamma rays may also result from loop-induced 
annihilations \cite{gammaline,zgamma}
$\chi\chi\to\gamma\gamma$ or $\chi\chi\to Z\gamma$.

This would give monoenergetic photons with $E_\gamma = m_\chi$ or $E_\gamma = m_\chi
(1-m_{Z}^2/4m_{\chi}^2)$ from the halo.
 The detection probability of a gamma line signal depends on
the very poorly known density profile of the dark matter halo.
N-body simulations have given a clue to the final halo 
profile obtained by hierarchical clustering in a CDM scenario \cite{NFW}. It turns out that the universal halo profile found in 
these simulations has a rather significant enhancement $\propto 1/r$ 
near the halo centre. If applicable to the Milky Way, this
 would lead to a much enhanced annihilation 
rate towards the galactic centre, and also to a very characteristic 
angular dependence of the line (and continuous gamma) signal. This would be very beneficial 
when discriminating against the galactic and extragalactic $\gamma$ 
ray background, and Air Cherenkov Telescopes (ACTs) would be eminently 
suited to look for these signals, if the energy resolution is at the 
$10-20$ \% level.

The models which give 
the highest rates should be within reach of the new generation of ACTs 
presently being constructed. These will have an effective area of almost $10^5$
m$^2$, a threshold of some tens of GeV and an energy resolution 
approaching 10 \%. For low-mass models, the space-borne
telescope GLAST will have a better sensitivity. (See \cite{BBU} for details.)

Another possibility to detect dark matter in gamma-rays has recently been
investigated \cite{diffuse2}. If N-body simulations of structure 
formation are taken seriously, it appears that the average enhancement
of the integrated signal from all cosmic structure in the universe would
be several orders of magnitude compared to the case when the dark matter
density only scales with the cosmic dilution factor $(1+z)^3$. 
The signature would be a continuum from neutralino annihilations plus a 
characteristic redshift-smeared line with a very rapid fall-off beyond
the energy corresponding to the neutralino mass. 

\subsection{Indirect detection through neutrinos}
More model-independent predictions, where essentially only the relatively
well-determined local halo dark matter density is of importance,
can be made for neutrinos  from the centre
of the Sun or Earth, where
neutralinos may have been gravitationally trapped and therefore their density
enhanced. (However, also the less well known velocity distribution also
does affect the capture rate, weakening the predictions somewhat, especially
for the Earth.)
 As neutralinos annihilate, many of the possible final states
give
after decays and perhaps hadronization energetic neutrinos which
will propagate out from the interior of the Sun or Earth.
In particular,
the muon neutrinos are useful for indirect detection of
neutralino annihilation processes, since muons
have a quite long range in a suitable detector medium like ice or water.
Therefore they can be detected through their Cherenkov radiation after
having been produced
at or near the detector, through the action of
a charged current weak interaction
$\nu_\mu + A \to \mu + X$.

Detection of neutralino annihilation into neutrinos is
one of the most promising indirect detection methods,
and  will be subject to  extensive experimental investigations in view
of the new neutrino telescopes (AMANDA/IceCUBE, Baikal, NESTOR, ANTARES)
planned or under construction \cite{halzenreview}. The advantage shared with
gamma rays is that neutrinos keep their original direction.

The detector threshold energy, for
``small''
neutrino telescopes like Baksan, MACRO (now discontinued) and 
Super-Kamiokande is
around 1 GeV.
Large area neutrino telescopes in the ocean  or in Antarctic ice
typically
have thresholds of the order of tens of GeV, which makes them
sensitive mainly to heavy neutralinos (above 100 GeV) \cite{begnu2}. 

The detector response is weighted towards high
neutrino energies, both because the cross section $\sigma_\nu$
rises approximately linearly with energy and because the average
muon energy, and therefore the range, also grow
approximately linearly with $E_\nu$. Therefore, final states
which give a hard neutrino spectrum (such as heavy quarks, $\tau$
leptons and $W$ or $Z$ bosons) are usually more important
than the soft spectrum arising from light quarks and gluons.

The capture rate, and therefore the signal rate, of the 
Earth is very 
similar to the scattering rate in the materials used for
spin-independent detection. It seems therefore that the new
generation of direct detection experiments generally have a greater 
sensitivity than the new neutrino telescopes searching for 
signals from the Earth \cite{kamsad}. For the Sun, however, neutrinos provide
a complementary and in some cases superior method \cite{lbreview}.  

\section{\uppercase{ Non-SUSY candidates}}

The phenomenology of supersymmetric dark matter (neutralinos) may be very
similar for other types of weakly interacting massive particles (WIMPs).
However, one can also imagine models where the WIMP only couples to
leptons. These leptonic WIMPs, or LIMPs, may at first seem essentially
undetectable in present-day experiments. It may be shown, however,
that in most cases they necessarily give energetic gamma rays in their
annihilations, due to higher-order processes \cite{limp}.

Of course, the current interest among theoretical physicists in exotic,
but at least partly realistic models like branes and large extra dimensions
has also led to a number of dark matter models. Just as an example,
models where the Standard Model fields live in
compactified, ``universal'' extra dimension has a rich dark matter
phenomenology, where an orbifold parity plays the role of R-parity 
in supersymmetric models. Detection rates turn out to be similar 
to the SUSY case but some interesting differences in the detailed 
phenomenology \cite{ued}.

It finally should be noted with caution, that there are many models
for dark matter where the candidate only interacts very weakly,
e.g.\,~gravitationally, with other matter. For some models, 
there still may exist cosmological
signals from early decays \cite{swimp}, but it may not be excluded that
 detection by other
methods is impossible. Still one may hope to learn more about
the detailed distribution of dark matter, for instance by studying
their effects on gravitational lensing. There are in fact studies \cite{lens}
that interpret flux ratios of multiple gravitational images in terms of 
clumpiness of galactic halos, in rough agreement with what is found
in N-body simulations.   
\section{\uppercase{ Conclusions}}

To conclude,  non-baryonic cold dark matter is required to
explain new cosmological data, in particular on the microwave background.
The fact that the favoured value of $\Omega_M$ has gone down 
during the last decade from near 1 to around 0.3 is good news for detection, 
since larger cross sections
generally means lower relic density. In particular this is true for the main
particle physics candidate, the neutralino, which we have presented
in some detail here. Direct detection experiments are rapidly developing
their discovery potential. Indirect detection methods may
be very useful complements.  
In particular, new gamma-ray and neutrino telescopes
may have the sensitivity to rule out or confirm the supersymmetry 
solution of the dark matter problem. 
And of course, when the LHC starts taking data, we may get direct clues 
to the nature of the particle that most likely makes up the dark matter.

\section{\uppercase{ Acknowledgements}}
I wish to thank my collaborators, in particular Ted Baltz, Joakim Edsj\"o, Paolo Gondolo, 
Mia Schelke and
Piero Ullio, for many
helpful discussions. This work has been supported in part by the
Swedish Research Council (VR).

\end{document}